# Game Theory with Translucent Players

Joseph Y. Halpern*
Cornell University
Dept. Computer Science
Ithaca, NY 14853, USA
halpern@cs.cornell.edu

Rafael Pass[†]
Cornell University
Dept. Computer Science
Ithaca, NY 14853, USA
rafael@cs.cornell.edu

## ABSTRACT

A traditional assumption in game theory is that players are opaque to one another—if a player changes strategies, then this change in strategies does not affect the choice of other players' strategies. In many situations this is an unrealistic assumption. We develop a framework for reasoning about games where the players may be *translucent* to one another; in particular, a player may believe that if she were to change strategies, then the other player would also change strategies. Translucent players may achieve significantly more efficient outcomes than opaque ones.

Our main result is a characterization of strategies consistent with appropriate analogues of common belief of rationality. *Common Counterfactual Belief of Rationality (CCBR)* holds if (1) everyone is rational, (2) everyone counterfactually believes that everyone else is rational (i.e., all players $i$ believe that everyone else would still be rational even if $i$ were to switch strategies), (3) everyone counterfactually believes that everyone else is rational, and counterfactually believes that everyone else is rational, and so on. CCBR characterizes the set of strategies surviving iterated removal of *minimax dominated* strategies: a strategy $\sigma_i$ is minimax dominated for $i$ if there exists a strategy $\sigma'_i$ for $i$ such that $\min_{\mu'_{-i}} u_i(\sigma'_i, \mu'_{-i}) > \max_{\mu_{-i}} u_i(\sigma_i, \mu_{-i})$.

## Categories and Subject Descriptors

F.4.1 [**Mathematical Logic and Formal Languages**]: Mathematical Logic—*modal logic*; I.2.11 [**Artificial Intelligence**]: Distributed Artificial Intelligence—*multiagent systems*; J.4 [**Social and Behavioral Sciences**]: Economics

## 1. INTRODUCTION

Two large firms 1 and 2 need to decide whether to *cooperate (C)* or *sue (S)* the other firm. Suing the other firm always has a small positive reward, but being sued induces a high penalty $p$; more precisely, $u(C,C) = (0,0); u(C,S) = (-p,r); u(S,C) = (r,-p), u(S,S) = (r-p, r-p)$. In other words, we are considering an instance of the Prisoner's Dilemma.

But there is a catch. Before acting, each firms needs to discuss their decision with its board. Although these discussions are held behind closed doors, there is always the possibility of the decision being "leaked"; as a consequence, the other company may change its course of action. Furthermore, both companies are aware of this fact. In other words, the players are *translucent* to one another.

In such a scenario, it may well be rational for both companies to cooperate. For instance, consider the following situation.

- Firm $i$ believes that its action is leaked to firm $2-i$ with probability $\epsilon$.

- Firm $i$ believes that if the other firm $2-i$ finds out that $i$ is defecting, then $2-i$ will also defect.

- Finally, $p\epsilon > r$ (i.e., the penalty for being sued is significantly higher than the reward of suing the other company).

Neither firm defects, since defection is noticed by the other firm with probability $\epsilon$, which (according to their beliefs) leads to a harsh punishment. Thus, the possibility of the players' actions being leaked to the other player allows the players to significantly improve social welfare in equilibrium. (This suggests that it may be mutually beneficial for two countries to spy on each other!)

Even if the Prisoner's dilemma is not played by corporations but by individuals, each player may believe that if he chooses to defect, his "guilt" over defecting may be visible to the other player. (Indeed, facial and bodily cues such as increased pupil size are often associated with deception; see e.g., [Ekman and Friesen 1969].) Thus, again, the players may choose to cooperate out of fear that if they defect, the other player may detect it and act on it.

Our goal is to capture this type of reasoning formally. We take a Bayesian approach: Each player has a (subjective) probability distribution (describing the player's beliefs) over the states of the world. Traditionally, a player $i$ is said to be rational in a state $\omega$ if the strategy $\sigma_i$ that $i$ plays at $\omega$ is a best response to the strategy profile $\mu_{-i}$ of the other players induced by $i$'s beliefs in $\omega$;[1] that is,

---
[1]Formally, we assume that $i$ has a distribution on states, and at each

*Halpern is supported in part by NSF grants IIS-0812045, IIS-0911036, and CCF-1214844, by AFOSR grant FA9550-08-1-0266, and by ARO grant W911NF-09-1-0281.

[†]Pass is supported in part by a Alfred P. Sloan Fellowship, Microsoft New Faculty Fellowship, NSF Award CNS-1217821, NSF CAREER Award CCF-0746990, NSF Award CCF-1214844, AFOSR YIP Award FA9550-10-1-0093, and DARPA and AFRL under contract FA8750-11-2- 0211. The views and conclusions contained in this document are those of the authors and should not be interpreted as representing the official policies, either expressed or implied, of the Defense Advanced Research Projects Agency or the US Government

$u_i(\sigma_i, \mu_{-i}) \geq u_i(\sigma'_i, \mu_{-i})$ for all alternative strategies $\sigma'_i$ for $i$. In our setting, things are more subtle. Player $i$ may believe that if she were to switch strategies from $\sigma_i$ to $\sigma'_i$, then players other than $i$ might also switch strategies. We capture this using *counterfactuals* [Lewis 1973; Stalnaker 1968].[2] Associated with each state of the world $\omega$, each player $i$, and $f(\omega, i, \sigma'_i)$ where player $i$ plays $\sigma'_i$. Note that if $i$ changes strategies, then this change in strategies may start a chain reaction, leading to further changes. We can think of $f(\omega, i, \sigma'_i)$ as the steady-state outcome of this process: the state that would result if $i$ switched strategies to $\sigma'_i$. Let $\mu_{f(\omega,i,\sigma'_i)}$ be the distribution on strategy profiles of $-i$ (the players other than $i$) induced by $i$'s beliefs at $\omega$ about the steady-state outcome of this process. We say that $i$ is rational at a state $\omega$ where $i$ plays $\sigma_i$ and has beliefs $\mu_i$ if $u_i(\sigma_i, \mu_{-i}) \geq u_i(\sigma'_i, \mu_{f(\omega,i,\sigma'_i)})$ for every alternative strategy $\sigma'_i$ for $i$. Note that we have required the closest-state function to be deterministic, returning a unique state, rather than a distribution over states. While this may seem incompatible with the motivating scenario, it does not seem so implausible in our context that, by taking a rich enough representation of states, we can assume that a state contains enough information about players to resolve uncertainty about what strategies they would use if one player were to switch strategies.

We are interested in considering analogues to rationalizability in a setting with translucent players, and providing epistemic characterizations of them. To do that, we need some definitions. We say that a player $i$ *counterfactually believes* $\varphi$ at $\omega$ if $i$ believes $\varphi$ holds even if $i$ were to switch strategies. *Common Counterfactual Belief of Rationality (CCBR)* holds if (1) everyone is rational, (2) everyone counterfactually believes that everyone else is rational (i.e., all players $i$ believe that everyone else would still be still rational even if $i$ were to switch strategies), (3) everyone counterfactually believes that everyone else is rational, and counterfactually believes that everyone else is rational, and so on.

Our main result is a characterization of strategies consistent with CCBR. Roughly speaking, these results can be summarized as follows:

- If the closest-state function respects "unilateral deviations"—when $i$ switches strategies, the strategies and beliefs of players other than $i$ remain the same—then CCBR characterizes the set of rationalizable strategies.

- If the closest-state function can be arbitrary, CCBR char-

---

state, a pure strategy profile is played; the distribution on states clearly induces a distribution on strategy profiles for the players other than $i$, which we denote $\mu_{-i}$.

[2] A different, more direct, approach for capturing our original motivating example would be to consider and analyze an extensive-form variant $G'$ of the original normal-form game $G$ that explicitly models the "leakage" of players' actions in $G$, allows the player to react to these leakage signals by choosing a new action in $G$, which again may be leaked and the players may react to, and so on. Doing this is subtle. We would need to model how players respond to receiving leaked information, and to believing that there was a change in plan even if information wasn't leaked. To make matters worse, it's not clear what it would mean that a player is "intending" to perform an action $a$ if players can revise what they do as the result of a leak. Does it mean that a player will do $a$ if no information is leaked to him? What if no information is leaked, but he believes that the other side is planning to change their plans in any case? In addition, modeling the game in this way would require a distribution over leakage signals to be exogenously given (as part of the description of the game $G'$). Moreover, player strategies would have to be infinite objects, since there is no bound on the sequence of leaks and responses to leaks. In contrast, using counterfactuals, we can directly reason about the original (finite) game $G$.

acterizes the set of strategies that survive iterated removal of *minimax dominated* strategies: a strategy $\sigma_i$ is minimax dominated for $i$ if there exists a strategy $\sigma'_i$ for $i$ such that $\min_{\mu'_{-i}} u_i(\sigma'_i, \mu'_{-i}) > \max_{\mu_{-i}} u_i(\sigma_i, \mu_{-i})$; that is, $u_i(\sigma'_i, \mu'_{-i}) > u_i(\sigma_i, \mu_{-i})$ no matter what the strategy profiles $\mu_{-i}$ and $\mu'_{-i}$ are.

We also consider analogues of Nash equilibrium in our setting, and show that individually rational strategy profiles that survive iterated removal of minimax dominated strategies characterize such equilibria.

Note that in our approach, each player $i$ has a *belief* about how the other players' strategies would change if $i$ were to change strategies, but we do not require $i$ to explicitly specify how he would respond to other people changing strategies. The latter approach, of having each player pick a "meta-strategy" that takes as input the strategy of other players, was explored by Howard [1971] in the 1970s. It led to complex formalisms involving infinite hierarchies of meta-strategies: at the lowest level, each player specifies a strategy in the original game; at level $k$, each player specifies a "response rule" (i.e., a meta-strategy) to other players' $(k-1)$-level response rules. Such hierarchical structures have not proven useful when dealing with applications. Since we do not require players to specify reaction rules, we avoid the complexities of this approach.

*Program equilibria* [Tennenholz 2004] and *conditional commitments* [Kalai et al. 2010] provide a different approach to avoiding infinite hierarchies. Roughly speaking, each player $i$ simply specifies a *program* $\Pi_i$; player $i$'s action is determined by running $i$'s program on input the (description of) the programs of the other players; that is, $i'$ action is given by $\Pi_i(\Pi_{-i})$. Tennenholtz [2004] and Kalai et al. [2010] show that every (correlated) individually rational outcome can be sustained in a program equilibrium. Their model, however, assumes that player's programs (which should be interpreted as their "plan of action") are commonly known to all players. We dispense with this assumption. It is also not clear how to define common belief of rationality in their model; the study of program equilibria and conditional commitments has considered only analogues of Nash equilibrium.

Counterfactuals have been explored in a game-theoretic setting; see, for example, [Aumann 1995; Halpern 1999; Samet 1996; Stalnaker 1996; Zambrano 2004]. However, all these papers considered only structures where, in the closest state where $i$ changes strategies, all other players' strategies remain the same; thus, these approaches are not applicable in our context.

## 2. COUNTERFACTUAL STRUCTURES

Given a game $\Gamma$, let $\Sigma_i(\Gamma)$ denote player $i$'s pure strategies in $\Gamma$ (we occasionally omit the parenthetical $\Gamma$ if it is clear from context or irrelevant).

To reason about the game $\Gamma$, we consider a class of Kripke structures corresponding to $\Gamma$. For simplicity, we here focus on finite structures. A *finite probability structure $M$ appropriate for* $\Gamma$ is a tuple $(\Omega, \mathbf{s}, \mathcal{PR}_1, \ldots, \mathcal{PR}_n)$, where $\Omega$ is a finite set of states; $\mathbf{s}$ associates with each state $\omega \in \Omega$ a pure strategy profile $\mathbf{s}(\omega)$ in the game $\Gamma$; and, for each player $i$, $\mathcal{PR}_i$ is a *probability assignment* that associates with each state $\omega \in \Omega$ a probability distribution $\mathcal{PR}_i(\omega)$ on $\Omega$, such that

1. $\mathcal{PR}_i(\omega)(\llbracket \mathbf{s}_i(\omega) \rrbracket_M) = 1$, where for each strategy $\sigma_i$ for player $i$, $\llbracket \sigma_i \rrbracket_M = \{\omega : \mathbf{s}_i(\omega) = \sigma_i\}$, where $\mathbf{s}_i(\omega)$ denotes player $i$'s strategy in the strategy profile $\mathbf{s}(\omega)$;

2. $\mathcal{PR}_i(\omega)(\llbracket \mathcal{PR}_i(\omega), i \rrbracket_M) = 1$, where for each probability measure $\pi$ and player $i$, $\llbracket \pi, i \rrbracket_M = \{\omega : \mathcal{PR}_i(\omega) = \pi\}$.

These assumptions say that player $i$ assigns probability 1 to his actual strategy and beliefs.

To deal with counterfactuals, we augment probability structures with a "closest-state" function $f$ that associates with each state $\omega$, player $i$, and strategy $\sigma'_i$, a state $f(\omega, i, \sigma_i)$ where player $i$ plays $\sigma'$; if $\sigma'$ is already played in $\omega$, then the closest state to $\omega$ where $\sigma'$ is played is $\omega$ itself. Formally, a *finite counterfactual structure* $M$ appropriate for $\Gamma$ is a tuple $(\Omega, \mathbf{s}, f, \mathcal{PR}_1, \ldots, \mathcal{PR}_n)$, where $(\Omega, \mathbf{s}, \mathcal{PR}_1, \ldots, \mathcal{PR}_n)$ is a probability structure appropriate for $\Gamma$ and $f$ is a "closest-state" function. We require that if $f(\omega, i, \sigma'_i) = \omega'$, then

1. $\mathbf{s}_i(\omega') = \sigma'$;
2. if $\sigma'_i = \mathbf{s}_i(\omega)$, then $\omega' = \omega$.

Given a probability assignment $\mathcal{PR}_i$ for player $i$, we define $i$'s counterfactual belief at state $\omega$ ("what $i$ believes would happen if he switched to $\sigma'_i$ at $\omega$") as

$$\mathcal{PR}^c_{i,\sigma'_i}(\omega)(\omega') = \sum_{\{\omega'' \in \Omega : f(\omega'', i, \sigma'_i) = \omega'\}} \mathcal{PR}_i(\omega)(\omega'').$$

Note that the conditions above imply that each player $i$ knows what strategy he would play if he were to switch; that is, $\mathcal{PR}^c_{i,\sigma'_i}(\omega)(\llbracket \sigma'_i \rrbracket_M) = 1$.

Let $Supp(\pi)$ denote the support of the probability measure $\pi$. Note that $Supp(\mathcal{PR}^c_{i,\sigma'_i}(\omega)) = \{f(\omega', i, \sigma'_i) : \omega' \in Supp(\mathcal{PR}_i(\omega))\}$. Moreover, it is almost immediate from the definition that if $\mathcal{PR}_i(\omega) = \mathcal{PR}_i(\omega')$, then $\mathcal{PR}^c_{i,\sigma'_i}(\omega) = \mathcal{PR}^c_{i,\sigma'_i}(\omega')$ for all strategies $\sigma'_i$ for player $i$. But it does *not* in general follow that $i$ knows his counterfactual beliefs at $\omega$, that is, it may not be the case that for all strategies $\sigma'_i$ for player $i$, $\mathcal{PR}^c_{i,\sigma'_i}(\omega)(\llbracket \mathcal{PR}^c_{i,\sigma'_i}(\omega), i \rrbracket_M) = 1$. Suppose that we think of a state as representing each player's *ex ante* view of the game. The fact that player $\mathbf{s}_i(\omega) = \sigma_i$ should then be interpreted as "$i$ intends to play $\sigma_i$ at state $\omega$." With this view, suppose that $\omega$ is a state where $\mathbf{s}_i(\omega)$ is a conservative strategy, while $\sigma'_i$ is a rather reckless strategy. It seems reasonable to expect that $i$'s subjective beliefs regarding the likelihood of various outcomes may depend in part on whether he is in a conservative or reckless frame of mind. We can think of $\mathcal{PR}^c_{i,\sigma'_i}(\omega)(\omega')$ as the probability that $i$ ascribes, at state $\omega$, to $\omega'$ being the outcome of $i$ switching to strategy $\sigma'_i$; thus, $\mathcal{PR}^c_{i,\sigma'_i}(\omega)(\omega')$ represents $i$'s evaluation of the likelihood of $\omega'$ when he is in a conservative frame of mind. This may not be the evaluation that $i$ uses in states in the support $\mathcal{PR}^c_{i,\sigma'_i}(\omega)$; at all these states, $i$ is in a "reckless" frame of mind. Moreover, there may not be a unique reckless frame of mind, so that $i$ may not have the same beliefs at all the states in the support of $\mathcal{PR}^c_{i,\sigma'_i}(\omega)$.

$M$ is a *strongly appropriate counterfactual structure* if it is an appropriate counterfactual structure and, at every state $\omega$, every player $i$ knows his counterfactual beliefs. As the example above suggests, strong appropriateness is a nontrivial requirement. As we shall see, however, our characterization results hold in both appropriate and strongly appropriate counterfactual structures.

Note that even in strongly appropriate counterfactually structures, we may not have $\mathcal{PR}_i(f(\omega, i, \sigma'_i)) = \mathcal{PR}^c_{i,\sigma'_i}(\omega)$. We do have $\mathcal{PR}_i(f(\omega, i, \sigma'_i)) = \mathcal{PR}^c_{i,\sigma'_i}(\omega)$ in strongly appropriate counterfactual structures if $f(\omega, i, \sigma'_i)$ is in the support of $\mathcal{PR}^c_{i,\sigma'_i}(\omega)$ (which will certainly be the case if $\omega$ is in the support of $\mathcal{PR}_i(\omega)$). To see why we may not want to have $\mathcal{PR}_i(f(\omega, i, \sigma'_i)) = \mathcal{PR}^c_{i,\sigma'_i}(\omega)$ in general, even in strongly appropriate counterfactual structures, consider the example above again. Suppose that, in state $\omega$, although $i$ does not realize it, he has been given a drug that affects how he evaluates the state. He thus ascribes probability 0 to $\omega$. In $f(\omega, i, \sigma'_i)$ he has also been given the drug, and the drug in particular affects how he evaluates outcomes. Thus, $i$'s beliefs in the state $f(\omega, i, \sigma'_i)$ are quite different from his beliefs in all states in the support of $\mathcal{PR}^c_{i,\sigma'_i}(\omega)$.

### 2.1 Logics for Counterfactual Games

Let $\mathcal{L}(\Gamma)$ be the language where we start with *true* and the primitive proposition $RAT_i$ and $play_i(\sigma_i)$ for $\sigma_i \in \Sigma_i(\Gamma)$, and close off under the modal operators $B_i$ (player $i$ believes) and $B^*_i$ (player $i$ counterfactually believes) for $i = 1, \ldots, n$, $CB$ (common belief), and $CB^*$ (common counterfactual belief), conjunction, and negation. We think of $B_i\varphi$ as saying that "$i$ believes $\varphi$ holds with probability 1" and $B^*_i\varphi$ as saying "$i$ believes that $\varphi$ holds with probability 1, even if $i$ were to switch strategies".

Let $\mathcal{L}^0$ be defined exactly like $\mathcal{L}$ except that we exclude the "counterfactual" modal operators $B^*$ and $CB^*$. We first define semantics for $\mathcal{L}^0$ using probability structures (without counterfactuals). We define the notion of a formula $\varphi$ being true at a state $\omega$ in a probability structure $M$ (written $(M, w) \models \varphi$) in the standard way, by induction on the structure of $\varphi$, as follows:

- $(M, \omega) \models true$ (so *true* is vacuously true).

- $(M, \omega) \models play_i(\sigma_i)$ iff $\sigma_i = \mathbf{s}_i(\omega)$.

- $(M, \omega) \models \neg\varphi$ iff $(M, \omega) \not\models \varphi$.

- $(M, \omega) \models \varphi \wedge \varphi'$ iff $(M, \omega) \models \varphi$ and $(M, \omega) \models \varphi'$.

- $(M, \omega) \models B_i\varphi$ iff $\mathcal{PR}_i(\omega)(\llbracket \varphi \rrbracket_M) = 1$, where $\llbracket \varphi \rrbracket_M = \{\omega : (M, \omega) \models \varphi\}$.

- $(M, \omega) \models RAT_i$ iff $\mathbf{s}_i(\omega)$ is a best response given player $i$'s beliefs regarding the strategies of other players induced by $\mathcal{PR}_i$.

- Let $EB\varphi$ ("everyone believes $\varphi$") be an abbreviation of $B_1\varphi \wedge \ldots \wedge B_n\varphi$; and define $EB^k\varphi$ for all $k$ inductively, by taking $EB^1\varphi$ to be $EB\varphi$ and $EB^{k+1}\varphi$ to be $EB(EB^k\varphi)$.

- $(M, \omega) \models CB\varphi$ iff $(M, \omega) \models EB^k\varphi$ for all $k \geq 1$.

Semantics for $\mathcal{L}^0$ in counterfactual structures is defined in an identical way, except that we redefine $RAT_i$ to take into account the fact that player $i$'s beliefs about the strategies of players $-i$ may change if $i$ changes strategies.

- $(M, \omega) \models RAT_i$ iff for every strategy $\sigma'_i$ for player $i$,

$$\sum_{\omega' \in \Omega} \mathcal{PR}_i(\omega)(\omega')u_i(\mathbf{s}_i(\omega), \mathbf{s}_{-i}(\omega')) \geq$$

$$\sum_{\omega' \in \Omega} \mathcal{PR}^c_{i,\sigma'_i}(\omega)(\omega')u_i(\sigma'_i, \mathbf{s}_{-i}(\omega')).$$

The condition above is equivalent to requiring that

$$\sum_{\omega' \in \Omega} \mathcal{PR}_i(\omega)(\omega')u_i(\mathbf{s}_i(\omega), \mathbf{s}_{-i}(\omega')) \geq$$

$$\sum_{\omega' \in \Omega} \mathcal{PR}_i(\omega)(\omega')u_i(\sigma'_i, \mathbf{s}_{-i}(f(\omega', i, \sigma'_i))).$$

Note that, in general, this condition is different from requiring that $\mathbf{s}_i(\omega)$ is a best reponse given player $i$'s beliefs regarding the strategies of other players induced by $\mathcal{PR}_i$.

To give the semantics for $\mathcal{L}$ in counterfactual structures, we now also need to define the semantics of $B^*_i$ and $CB^*$:

- $(M, \omega) \models B_i^* \varphi$ iff for all strategies $\sigma_i' \in \Sigma_i(\Gamma)$, $\mathcal{PR}_{i,\sigma_i'}^c(\omega)(\llbracket\varphi\rrbracket_M) = 1$.

- $(M, \omega) \models CB^* \varphi$ iff $(M, \omega) \models (EB^*)^k \varphi$ for all $k \geq 1$.

It is easy to see that, like $B_i$, $B_i^*$ depends only on $i$'s beliefs; as we observed above, if $\mathcal{PR}_i(\omega) = \mathcal{PR}_i(\omega')$, then $\mathcal{PR}_{i,\sigma_i'}^c(\omega) = \mathcal{PR}_{i,\sigma_i'}^c(\omega')$ for all $\sigma_i'$, so $(M, \omega) \models B_i^* \varphi$ iff $(M, \omega') \models B_i^* \varphi$. It immediately follows that $B_i^* \varphi \Rightarrow B_i B_i^* \varphi$ is valid (i.e., true at all states in all structures).

The following abbreviations will be useful in the sequel. Let $RAT$ be an abbreviation for $RAT_1 \wedge \ldots \wedge RAT_n$, and let $play(\vec{\sigma})$ be an abbreviation for $play_1(\sigma_1) \wedge \ldots \wedge play_n(\sigma_n)$.

## 2.2 Common Counterfactual Belief of Rationality

We are interested in analyzing strategies being played at states where (1) everyone is rational, (2) everyone counterfactually believes that everyone else is rational (i.e., for every player $i$, $i$ believes that everyone else would still be rational even if $i$ were to switch strategies), (3) everyone counterfactually believes that everyone else is rational, and counterfactually believes that everyone else is rational, and so on. For each player $i$, define the formulas $SRAT_i^k$ (player $i$ is strongly $k$-level rational) inductively, by taking $SRAT_i^0$ to be $true$ and $SRAT_i^{k+1}$ to be an abbreviation of

$$RAT_i \wedge B_i^*(\wedge_{j \neq i} SRAT_j^k).$$

Let $SRAT^k$ be an abbreviation of $\wedge_{j=1}^n SRAT_j^k$.

Define $CCBR$ (common counterfactual belief of rationality) as follows:

- $(M, \omega) \models CCBR$ iff $(M, \omega) \models SRAT^k \varphi$ for all $k \geq 1$.

Note that it is critical in the definition of $SRAT_i^k$ that we require only that player $i$ counterfactually believes that everyone else (i.e., the players other than $i$) are rational, and believe that everyone else is rational, and so on. Player $i$ has no reason to believe that his own strategy would be rational if he were to switch strategies; indeed, $B_i^* RAT_i$ can hold only if *every* strategy for player $i$ is rational with respect to $i$'s beliefs. This is why we do not define $CCBR$ as $CB^*RAT$.[3]

We also consider the consequence of just common belief of rationality in our setting. Define $WRAT_i^k$ (player $i$ is weakly $k$-level rational) just as $SRAT_i^k$, except that $B_i^*$ is replaced by $B_i$. An easy induction on $k$ shows that $WRAT^{k+1}$ implies $WRAT^k$ and that $WRAT^k$ implies $B_i(WRAT^k)$.[4] It follows that we could have equivalently defined $WRAT_i^{k+1}$ as

$$RAT_i \wedge B_i(\wedge_{j=1}^n WRAT_j^k).$$

Thus, $WRAT^{k+1}$ is equivalent to $RAT \wedge EB(WRAT^k)$. As a consequence we have the following:

PROPOSITION 2.1: $(M, \omega) \models CB(RAT)$ iff $(M, \omega) \models WRAT^k$ for all $k \geq 0$.

---

[3]Interestingly, Samet [1996] essentially considers an analogue of $CB^*RAT$. This works in his setting since he is considering only events in the past, not events in the future.

[4]We can also show that $SRAT^{k+1}$ implies $SRAT^k$, but it is not the case that $SRAT_i^k$ implies $B_i^* SRAT_i^k$, since $RAT$ does not imply $B_i^* RAT$.

## 3. CHARACTERIZING COMMON COUNTERFACTUAL BELIEF OF RATIONALITY

It is well known rationalizability can be characterized in terms of common belief of common belief of rationality in probability structures [**?**; **?**]. In the full version of the paper[5] we show that if we restrict to counterfactual structures that *respect unilateral deviations*—where in the closest state to $\omega$ where player $i$ switches strategies, everybody else's strategy and beliefs remain same—common counterfactual belief of rationality characterizes rationalizable strategies. In a sense (which is made precise in the full version of the paper), counterfactual structures respecting unilateral deviations behave just like probability structures (without counterfactuals).

We now characterize common counterfactual belief of rationality without putting any restrictions on the counterfactual structures (other than them being appropriate, or strongly appropriate). Our characterization is based on ideas that come from the characterization of rationalizability. It is well known that rationalizability can be characterized in terms of an iterated deletion procedure, where at each stage, a strategy $\sigma$ for player $i$ is deleted if there are no beliefs that $i$ could have about the undeleted strategies for the players other than $i$ that would make $\sigma$ rational [Pearce 1984]. Thus, there is a deletion procedure that, when applied repeatedly, results in only the rationalizable strategies, that is, the strategies that are played in states where there is common belief of rationality, being left undeleted. We now show that there is an analogous way of characterizing common counterfactual belief of rationality.

### 3.1 Iterated Minimax Domination

The key to our characterization is the notion of *minimax dominated* strategies.

DEFINITION 3.1: *Strategy $\sigma_i$ for player $i$ in game $\Gamma$ is* minimax dominated with respect to $\Sigma_{-i}' \subseteq \Sigma_{-i}(\Gamma)$ *iff there exists a strategy* $\sigma_i' \in \Sigma_i(\Gamma)$ *such that*

$$\min_{\tau_{-i} \in \Sigma_{-i}'} u_i(\sigma_i', \tau_{-i}) > \max_{\tau_{-i} \in \Sigma_{-i}'} u_i(\sigma_i, \tau_{-i}).$$

∎

In other words, player $i$'s strategy $\sigma$ is minimax dominated with respect to $\Sigma_{-i}'$ iff there exists a strategy $\sigma'$ such that the worst-case payoff for player $i$ if he uses $\sigma'$ is strictly better than his best-case payoff if he uses $\sigma$, given that the other players are restricted to using a strategy in $\Sigma_{-i}'$.

In the standard setting, if a strategy $\sigma_i$ for player $i$ is dominated by $\sigma_i'$ then we would expect that a rational player will never player $\sigma_i$, because $\sigma_i'$ is a strictly better choice. As is well known, if $\sigma_i$ is dominated by $\sigma_i'$, then there are no beliefs that $i$ could have regarding the strategies used by the other players according to which $\sigma_i$ is a best response [Pearce 1984]. This is no longer the case in our setting. For example, in the standard setting, cooperation is dominated by defection in Prisoner's Dilemma. But in our setting, suppose that player 1 believes that if he cooperates, then the other player will cooperate, while if he defects, then the other player will defect. Then cooperation is not dominated by defection.

So when can we guarantee that playing a strategy is irrational in our setting? This is the case only if the strategy is minimax dominated. If $\sigma_i$ is minimax dominated by $\sigma_i'$, there are no counterfactual beliefs that $i$ could have that would justify playing $\sigma_i$. Conversely, if $\sigma_i$ is not minimax dominated by any strategy, then there

---

[5]Available at http://www.cs.cornell.edu/home/halpern/papers/minimax.pdf.

219

are beliefs and counterfactual beliefs that $i$ could have that would justify playing $\sigma_i$. Specifically, $i$ could believe that the players in $-i$ are playing the strategy profile that gives $i$ the best possible utility when he plays $\sigma_i$, and that if he switches to another strategy $\sigma'_i$, the other players will play the strategy profile that gives $i$ the worst possible utility given that he is playing $\sigma'_i$.

Note that we consider only domination by pure strategies. It is easy to construct examples of strategies that are not minimax dominated by any pure strategy, but are minimax dominated by a mixed strategy. Our characterization works only if we restrict to domination by pure strategies. The characterization, just as with the characterization of rationalizability, involves iterated deletion, but now we do not delete dominated strategies in the standard sense, but minimax dominated strategies.

DEFINITION 3.2: *Define $NSD_j^k(\Gamma)$ inductively: let $NSD_j^0(\Gamma) = \Sigma_j$ and let $NSD_j^{k+1}(\Gamma)$ consist of the strategies in $NSD_j^k(\Gamma)$ not minimax dominated with respect to $NSD_{-j}^k(\Gamma)$. Strategy $\sigma$ survives $k$ rounds of iterated deletion of minimax strategies for player $i$ if $\sigma \in NSD_i^k(\Gamma)$. Strategy $\sigma$ for player $i$ survives iterated deletion of minimax dominated strategies if it survives $k$ rounds of iterated deletion of strongly dominated for all $k$, that is, if $\sigma \in NSD_i^\infty(\Gamma) = \cap_k NSD_i^k(\Gamma)$.* ∎

In the deletion procedure above, at each step we remove *all* strategies that are minimax dominated; that is we perform a "maximal" deletion at each step. As we now show, the set of strategies that survives iterated deletion is actually independent of the deletion order.

Let $S^0, \ldots, S^m$ be sets of strategy profiles. $\vec{S} = (S^0, S^1, \ldots, S^m)$ is a *terminating deletion sequence* for $\Gamma$ if, for $j = 0, \ldots, m-1$, $S^{j+1} \subset S^j$ (note that we use $\subset$ to mean proper subset) and all players $i$, $S_i^{j+1}$ contains all strategies for player $i$ not minimax dominated with respect to $S_{-i}^j$ (but may also contain some strategies that are minimax dominated), and $S_i^m$ does not contain any strategies that are minimax dominated with respect to $S_{-i}^m$. A set $T$ of strategy profiles has *ambiguous* terminating sets if there exist two terminating deletion sequences $\vec{S} = (T, S_1, \ldots, S_m)$, $\vec{S}' = (T, S'_1, \ldots, S'_{m'})$ such that $S_m \neq S'_{m'}$; otherwise, we say that $T$ has a *unique terminating set*.

PROPOSITION 3.3: *No (finite) set of strategy profiles has ambiguous terminating sets.*

**Proof:** Let $T$ be a set of strategy profiles of least cardinality that has ambiguous terminating deletion sequences $\vec{S} = (T, S_1, \ldots, S_m)$ and $\vec{S}' = (T, S'_1, \ldots, S'_{m'})$, where $S_m \neq S'_{m'}$. Let $T'$ be the set of strategies that are not minimax dominated with respect to $T$. Clearly $T' \neq \emptyset$ and, by definition, $T' \subseteq S_1 \cap S'_1$. Since $T'$, $S_1$, and $S'_1$ all have cardinality less than that of $T$, they must all have unique terminating sets; moreover, the terminating sets must be the same. For consider a terminating deletion sequence starting at $T'$. We can get a terminating deletion sequence starting at $S_1$ by just appending this sequence to $S_1$ (or taking this sequence itself, if $S_1 = T'$). We can similarly get a terminating deletion sequence starting at $S'_1$. Since all these terminating deletion sequences have the same final element, this must be the unique terminating set. But $(S_1, \ldots, S_m)$ and $(S'_1, \ldots, S'_{m'})$ are terminating deletion sequences with $S_m \neq S'_{m'}$, a contradiction. ∎

COROLLARY 3.4: *The set of strategies that survives interated deletion of minimax strategies is independent of the deletion order.*

REMARK 3.5: *Note that in the definition of $NSD_i^k(\Gamma)$, we remove all strategies that are dominated by some strategy in $\Sigma_i(\Gamma)$, not just those dominated by some strategy in $NSD_i^{k-1}(\Gamma)$. Nevertheless, the definition would be equivalent even if we had considered only dominating strategies in $NSD_i^{k-1}(\Gamma)$. For suppose not. Let $k$ be the smallest integer such that there exists some strategy $\sigma_i \in NSD_i^{k-1}(\Gamma)$ that is minimax dominated by a strategy $\sigma'_i \notin NSD_i^{k-1}(\Gamma)$, but there is no strategy in $NSD_i^{k-1}(\Gamma)$ that dominates $\sigma_i$. That is, $\sigma'_i$ was deleted in some previous iteration. Then there exists a sequence of strategies $\sigma_i^0, \ldots, \sigma_i^q$ and indices $k_0 < k_1 < \ldots < k_q = k - 1$ such that $\sigma_i^0 = \sigma_i$, $\sigma_i^j \in NSD_i^{k_j}(\Gamma)$, and for all $0 \leq j < q$, $\sigma_i^j$ is minimax dominated by $\sigma_i^{j+1}$ with respect to $NSD_i^{k_j - 1}(\Gamma)$. Since $NSD^{k-2}(\Gamma) \subseteq NSD^j(\Gamma)$ for $j \leq k-2$, an easy induction on $j$ shows that $\sigma_i^q$ minimax dominates $\sigma_i^{q-j}$ with respect to $NSD^{k-2}$ for all $0 < j \leq q$. In particular, $\sigma^q$ minimax dominates $\sigma_i^0 = \sigma'$ with respect to $NSD^{k-2}$.* ∎

The following example shows that iteration has bite: there exist a 2-player game where each player has $k$ actions and $k-1$ rounds of iterations are needed.

EXAMPLE 3.6: *Consider a two-player game, where both players announce a value between 1 and $k$. Both players receive in utility the smallest of the values announced; additionally, the player who announces the larger value get a reward of $p = 0.5$.[6] That is, $u(x, y) = (y + p, y)$ if $x > y$, $(x, x + p)$ if $y > x$, and $(x, x)$ if $x = y$. In the first step of the deletion process, 1 is deleted for both players; playing 1 can yield a max utility of 1, whereas the mininum utility of any other action is 1.5. Once 1 is deleted, 2 is deleted for both players: 2 can yield a max utility of 2, and the min utility of any other action (once 1 is deleted) is 2.5. Continuing this process, we see that only $(k, k)$ survives.* ∎

## 3.2 Characterizing Iterated Minimax Domination

We now show that strategies surviving iterated removal of minimax dominated strategies characterize the set of strategies consistent with common counterfactual belief of rationality in (strongly) appropriate counterfactual structures. As a first step, we define a "minimax" analogue of rationalizability.

DEFINITION 3.7: *A strategy profile $\vec{\sigma}$ in game $\Gamma$ is* minimax rationalizable *if, for each player $i$, there is a set $\mathcal{Z}_i \subseteq \Sigma_i(\Gamma)$ such that*

- *$\sigma_i \in \mathcal{Z}_i$;*

- *for every strategy $\sigma'_i \in \mathcal{Z}_i$ and strategy $\sigma''_i \in \Sigma_i(\Gamma)$,*

$$\max_{\tau_{-i} \in \mathcal{Z}_{-i}} u_i(\sigma'_i, \tau_{-i}) \geq \min_{\tau_{-i} \in \mathcal{Z}_{-i}} u_i(\sigma''_i, \tau_{-i}).$$

∎

THEOREM 3.8: *The following are equivalent:*

*(a) $\vec{\sigma} \in NSD^\infty(\Gamma)$;*

*(b) $\vec{\sigma}$ is minimax rationalizable in $\Gamma$;*

---

[6]This game can be viewed a a reverse variant of the Traveler's dilemma [Basu 1994], where the player who announces the smaller value gets the reward.

*(c) there exists a finite counterfactual structure $M$ that is strongly appropriate for $\Gamma$ and a state $\omega$ such that*

$$(M, \omega) \models play(\vec{\sigma}) \wedge \bigwedge_{i=1}^{n} SRAT_i^k$$

*for all $k \geq 0$;*

*(d) for all players $i$, there exists a finite counterfactual structure $M$ that is appropriate for $\Gamma$ and a state $\omega$ such that*

$$(M, \omega) \models play_i(\sigma_i) \wedge SRAT_i^k$$

*for all $k \geq 0$.*

The proof of Theorem 3.8 can be found in the full version of the paper. In the full version of the paper, we additionally characterize analogues of Nash equilibrium in counterfactual structures. These results allow us to more closely relate our model to those of Tennenholtz [2004] and Kalai et al. [2010].

## 4. DISCUSSION

We have introduced a game-theoretic framework for analyzing scenarios where a player may believe that if he were to switch strategies, this intention to switch may be detected by the other players, resulting in them also switching strategies. Our formal model allows players' counterfactual beliefs (i.e., their beliefs about the state of the world in the event that they switch strategies) to be arbitrary—they may be completely different from the players' actual beliefs.

We may also consider a more restricted model where we require that a player $i$'s counterfactual beliefs regarding other players' strategies and beliefs is $\epsilon$-close to player $i$'s actual beliefs in total variation distance[7]—that is, for every state $\omega \in \Omega$, player $i$, and strategy $\sigma_i'$ for player $i$, the projection of $\mathcal{PR}_{i,\sigma_i'}^c(\omega)$ onto strategies and beliefs of players $-i$ is $\epsilon$-close to the projection of $\mathcal{PR}_i(\omega)$ onto strategies and beliefs of players $-i$.

We refer to counterfactual structures satisfying this property as $\epsilon$-counterfactual stuctures. Roughly speaking, $\epsilon$-counterfactual structures restrict to scenarios where players are not "too" transparent to one another; this captures the situation when a player assigns only some "small" probability to its switch in action being noticed by the other players.

As we show in the full paper, 0-counterfactual structures behave just as counterfactual structures that respect unilateral deviations: common counterfactual belief of rationality in 0-counterfactual structures characterizes rationalizable strategies. The general counterfactual structures investigated in this paper are 1-counterfactual structures (that is, we do not impose any conditions on players' counterfactual beliefs). We remark that although our characterization results rely on the fact that we consider 1-counterfactual structures, the motivating example in the introduction (the translucent prisoner's dilemma game) shows that even considering $\epsilon$-counterfactual structures with a small $\epsilon$ can result in there being strategies consistent with common counterfactual belief of rationality that are not rationalizable. We leave an exploration of common counterfactual belief of rationality in $\epsilon$-counterfactual structures for future work.

## References

Aumann, R. J. (1995). Backwards induction and common knowledge of rationality. *Games and Economic Behavior 8*, 6–19.

Basu, K. (1994). The traveler's dilemma: paradoxes of rationality in game theory. *American Economic Review 84*(2), 391–395.

Brandenburger, A. and E. Dekel (1987). Rationalizability and correlated equilibria. *Econometrica 55*, 1391–1402.

Ekman, P. and W. Friesen (1969). Nonverbal leakage and clues to deception. *Psychiatry 32*, 88–105.

Halpern, J. Y. (1999). Hypothetical knowledge and counterfactual reasoning. *International Journal of Game Theory 28*(3), 315–330.

Howard, N. (1971). *Paradoxes of Rationality: Theory of Metagames and Political Behavior*. The MIT Press, Cambridge.

Kalai, A., E. Kalai, E. Lehrer, and D. Samet (2010). A commitment folk theorem. *Games and Economic Behavior 69*(1), 127–137.

Lewis, D. K. (1973). *Counterfactuals*. Cambridge, Mass.: Harvard University Press.

Pearce, D. G. (1984). Rationalizable strategic behavior and the problem of perfection. *Econometrica 52*(4), 1029–1050.

Samet, D. (1996). Hypothetical knowledge and games with perfect information. *Games and Economic Behavior 17*, 230–251.

Stalnaker, R. C. (1968). A semantic analysis of conditional logic. In N. Rescher (Ed.), *Studies in Logical Theory*, pp. 98–112. Oxford University Press.

Stalnaker, R. C. (1996). Knowledge, belief and counterfactual reasoning in games. *Economics and Philosophy 12*, 133–163.

Tan, T. and S. Werlang (1988). The Bayesian foundation of solution concepts of games. *Journal of Economic Theory 45*(45), 370–391.

Tennenholz, M. (2004). Program equilibrium. *Games and Economic Behavior 49*(12), 363–373.

Zambrano, E. (2004). Counterfactual reasoning and common knowledge of rationality in normal form. *Topics in Theoretical Economics 4*(1).

---

[7] Recall that two probability distribution are $\epsilon$-close in total variation distance if the probabilities that they assign to any event $E$ differ by at most $\epsilon$.